\def\h{{1\over 2}}

\documentstyle[amssymb]{article}
\begin{document}
{



\vspace{.2in}

\begin{center} {\LARGE NAHM'S EQUATIONS AND ROOT SYSTEMS}
\\ 
{\ }\\
Tomasz Brzezi\'nski\footnote{Research
supported by the EPSRC grant GR/K02244. On leave form: Department of
Theoretical Physics, University of \L\'od\'z, Pomorska 149/153, 90-236
\L\'od\'z, Poland} \&
Houari Merabet\footnote{Research
supported by the EPSRC grant GR/K50641}\\
 Department of Applied Mathematics \& Theoretical Physics\\
University of Cambridge, Cambridge CB3 9EW\\
 T.Brzezinski@damtp.cam.ac.uk\\
 M.Houari@damtp.cam.ac.uk\\
\end{center}
\begin{center}
\end{center}}

\vspace{10pt}
\begin{quote}
\noindent{\bf Abstract} A method of deriving solutions to Nahm's
equations based on root structure of simple Lie algebras is given. As
an illustration of this method the recently found solutions to
Nahm's  equations with tetrahedral and octahedral  symmetries are
shown to correspond to $A_2$ and $A_3$ root systems.
\bigskip

\end{quote}


{\bf 1.}
 It is well known \cite{Nahm:con} \cite{Hit:con}  that the
$N$-monopole solutions of the
self-dual $SU(2)$ Yang-Mills theory are equivalent to the Nahm
data. The latter consist of  meromorphic functions $T_1,T_2,T_3$
defined on the
 interval $[0,2]$, regular on  $(0,2)$ and  with values in $N\times N$
matrices. The $T_i$
satisfy  Nahm's equations
\begin{equation}
{d T_k(s)\over ds} ={1\over 2}\epsilon_{ijk} [T_i(s),T_j(s)],
\label{nahm.eqn}
\end{equation}
 and have simple poles at
0 and 2, the residues of which form an
irreducible $N$-dimensional representation of $su(2)$. Furthermore it
is required that the  $T_i$ satisfy the reality
conditions which, in a suitable basis, read:    $T_i\sp\dagger(s) =
-T_i(s)$ and $T_i(s) = T_i^t(2-s)$.

Recently,  solutions to the Nahm equations with tetrahedral,
octahedral and dodecahedral symmetries were found in \cite{HitMan:sym}
\cite{HouSut:oct}. These solutions were derived by requiring that
the monopole
spectral curve defined by the equation
$$
P(\eta,\zeta) \equiv \det (\eta + i(T_1 +iT_2)-2iT_3\zeta
-i(T_1-iT_2)\zeta^2) =0
$$
has the  regular solid symmetry. On the other hand it has been
known for some time
\cite{Rou:con} \cite{War:gen}
that Nahm's equations are closely related to the classical Yang-Baxter
equation which plays an important role  in the theory of integrable
models. Solutions to the
classical Yang-Baxter equation can be classified using the structure
of root spaces of simple Lie algebras \cite{BelDri:tri}. More recently
another class of equations resembling Nahm's equations appeared in
connection with integrable models of Calogero type. These are
integrability relations of the elliptic Dunkl differential operators
\cite{Dun:dif}. Classification of solutions to these equations given in
\cite{BucFel:ell} involves Weyl groups of
classical Lie algebras or, more generally, finite Coxeter groups. All
these recent developments in understanding equations appearing
naturally in the theory of integrable models strongly suggests that the proper
approach for solving Nahm's equations should involve root systems of simple Lie
algebras. In this paper we propose an ansatz for solving Nahm's
equations based on the root systems of $A_n$ type and we show that the
tetrahedral and octahedral solutions of \cite{HitMan:sym} fit into the
scheme. Starting with a few assumptions  we derive their basic
consequences concentrating on the Lie-algebraic
interpretation of existing solutions.
New solutions to Nahm's
equations together with detailed proofs will be presented elsewhere.
\medskip

{\bf 2.} Recall that any simple Lie algebra ${\cal L}$ of rank $r$
corresponding
to the root system $R$ is generated by $H_\mu$, $\mu = 1,\ldots , r$
and $E_\alpha$, $\alpha\in R$ which satisfy the following relations
$$
[H_\mu,H_\nu] = 0, \qquad [H_\mu, E_\alpha] = \alpha_\mu E_\alpha,
$$
$$
[E_\alpha, E_\beta] = N_{\alpha,\beta}E_{\alpha+\beta},\;\;
(\mbox{if $\alpha+\beta \in R$}), \quad [E_\alpha, E_{-\alpha}] =
\sum_{\mu =1}^r \alpha_\mu H_\mu,
$$
where $\alpha_\mu, N_{\alpha,\beta}$ are complex numbers. Since we are
interested in  solutions of a matrix equation we take any
$N$-dimensional representation of ${\cal L}$ chosen so that
$E^\dagger_{\alpha} =
E_{-\alpha}$ and $H_\mu^\dagger =H_\mu$. It is useful to represent
Nahm's data as a three-component vector field ${\bf T}(s)
= (T_1(s),T_2(s),T_3(s))$. Then Nahm's equations take the form ${d
\over ds}{\bf T}(s)  = {\bf T} (s)\wedge {\bf T} (s)$. We search
for solutions of this equation in the form
\begin{equation}
{\bf T}(s) = \sum_{\alpha\in R_+} ({\bf e}_\alpha(s) E_\alpha + {\bf
e}_{-\alpha}(s)  E_{-\alpha}),
\label{ansatz}
\end{equation}
where $R_+$ is a set of positive roots and ${\bf e}_\alpha(s)$ are
three-dimensional vector fields. The reality condition for the Nahm's
data $\bf T$ imply that ${\bf e}_{-\alpha} (s) = -{\bf e}_\alpha^*(s)$ while
(\ref{nahm.eqn}) becomes
\begin{equation}
{d{\bf e}_\beta(s)\over ds} = \h  \sum_{\alpha\in R}N_{\alpha,
\beta-\alpha} {\bf e}_\alpha(s)\wedge {\bf e}_{\beta-\alpha}(s)
\label{nahm.a.1}
\end{equation}
with a constraint
\begin{equation}
\sum_{\alpha\in R_+}\alpha_\mu {\bf e}_\alpha(s)\wedge {\bf
e}_{-\alpha}(s) = 0, \qquad \mu =1,\ldots, r.
\label{nahm.a.2}
\end{equation}
We are looking for the most symmetric configuration of monopoles,
i.e., we require that any exchange of at least two monopoles does
not change the configuration of the system. Motivated by the examples
discussed in section 3, we find that the maximal symmetry
requirement can be expressed as two conditions satisfied for all
positive roots $\alpha$
\begin{equation}
{\bf e}_\alpha(s)\cdot{\bf e}^*_\alpha(s) = f(s)
\label{as.1}
\end{equation}
and
\begin{equation}
{\bf e}_\alpha(s)\wedge{\bf e}^*_\alpha(s) =i g(s) {\bf v}_\alpha
\label{as.2}
\end{equation}
where $f(s)$ and $g(s)$ are real functions.
We assume that the function $g(s)$ does not depend
on the phases $\phi_{\alpha i}$, $i=1,2,3$,  of the components
of ${\bf e}_\alpha(s)$.
Equations (\ref{as.2}) combined with (\ref{nahm.a.2}) lead to
the following condition for the vectors ${\bf v}_\alpha$,
\begin{equation}
\sum_{\alpha\in R_+}\alpha_\mu {\bf v}_\alpha = 0, \qquad \mu =
1,2,\ldots , r.
\label{dep}
\end{equation}
Since the vectors ${\bf v}_\alpha$ are not linearly independent we impose
the following irreducibility and normalisation condition. The vector
${\bf v}_\alpha$ has a norm 1 if there are no vectors parallel to
${\bf v}_\alpha$ among
all the other ${\bf v}_\beta$. Otherwise we set ${\bf v}_\alpha$
to zero. Furthermore if ${\bf v}_\alpha = 0$ for all $\alpha$ in a subset
$P_+$ of $R_+$ and for $\beta \in R_+-P_+$,  ${\bf
e}_\beta(s)\wedge{\bf e}^*_\beta(s) \neq 0$, then ${\bf e}_\alpha(s)
=0$ for $\alpha\in P_+$. This last condition prevents a reduction of
the solution to Nahm's equations corresponding to a root system
of a Lie algebra of rank $r$  to the solution corresponding to a
Lie algebra of lower rank.

Combining (\ref{as.1}) and (\ref{as.2}) one obtains
$$
{\bf e}_\alpha(s)\cdot{\bf e}_\alpha(s) = |f(s)^2 -
g^2(s)|^{1/2}e^{i\theta_\alpha(s)} ,
$$
for some functions $\theta_\alpha$. On the other hand, taking
the derivative of (\ref{as.1}) and using (\ref{nahm.a.1})
one arrives at the following constraints
\begin{equation}
\sum_{\alpha\in R_+}N_{\alpha,
\beta-\alpha} ({\bf e}_\alpha(s)\wedge {\bf e}_{\beta-\alpha}(s))\cdot
{\bf e}_{\alpha+\beta}^*(s) + c.c = {d f(s)\over ds}
\label{constraint}
\end{equation}
In the $A_2$ case, this constraint is fulfilled automatically.
In the $A_3$ case,  the constraints (\ref{constraint}) are solved by
imposing  that
$N_{\alpha,\beta} ({\bf e}_\alpha(s)\wedge{\bf
e}_\beta(s))\cdot{\bf e}^*_{\alpha+\beta}(s)$
are real and  equal to each other for all $\alpha,\beta \in R$.
This gives
\begin{equation}
{d f(s)\over ds} = 4 N_{\alpha,\beta} ({\bf e}_\alpha(s)\wedge{\bf
e}_\beta(s))\cdot{\bf e}^*_{\alpha+\beta}(s),
\label{weier}
\end{equation}
\medskip

{\bf 3.} We now use the above ansatz to derive solutions to Nahm's
equations in the $A_2$ and $A_3$ cases. In the $A_2$ case equations
(\ref{nahm.a.1}) read
$$
{d {\bf e}_{\alpha_1}(s)\over ds} = - {\bf e}^*_{\alpha_2}(s)\wedge{\bf
e}_{\alpha_3}(s), \quad {d {\bf e}_{\alpha_2}(s)\over ds} =  {\bf
e}^*_{\alpha_1}(s)\wedge{\bf
e}_{\alpha_3}(s),
$$
\begin{equation}
 {d {\bf e}_{\alpha_3}(s)\over ds} = - {\bf
e}_{\alpha_1}(s)\wedge{\bf
e}_{\alpha_2}(s),
\label{a2.1}
\end{equation}
where $\alpha_1, \alpha_2$ are simple positive roots and $\alpha_3
=\alpha_1+\alpha_2$. The irreducibility condition together with
(\ref{dep}) imply that
${\bf v}_{\alpha_1} = {\bf v}_{\alpha_2} = {\bf v}_{\alpha_3} =
0$. This in turn determines the choice of the components of ${\bf
e}_{\alpha_i}(s)$ to be $e_{\alpha_i j} = e_j\delta_{ij}$ and reduces
equations (\ref{a2.1}) to
\begin{equation}
{d {e}_{1}(s)\over ds} = - {e}^*_{2}(s) e_{3}(s), \quad {d
{e}_{2}(s)\over ds} =  -{e}^*_{1}(s){
e}_{3}(s), \quad {d {e}_{3}(s)\over ds} = - {e}_{1}(s)
e_2(s).
\label{a2.2}
\end{equation}
The system of equations (\ref{a2.2}) can easily be solved using the
technique described in \cite{BucFel:ell}. One finds
$$
f(s) = {\cal P}(u), \qquad u = \zeta^{1/3} s +s_0,
$$
where $\zeta$ and $s_0$ are constants and $\cal P$ is the Weierstrass
elliptic function given as a solution of the equation  ${\cal P}'(u)^2
= 4({\cal
P}(u)^3-1)$. Setting $e_i(s) = f(s)^{1/2}e^{i\theta_i(s)}$
one finds that $\theta_2 = \theta_1+c_1$, $\theta_3 = -\theta_1 +c_2$,
and
\begin{equation}
\tan \left({1\over 2} \left(3\theta_1(s) +c_1-c_2\right)\right) = \pm
c_3 \exp({3\int f(s)^{1/2} ds}),
\label{gauge.a2}
\end{equation}
where $c_1, c_2, c_3$ are constants of integration. From the point of
view of the monopole dynamics a spectral curve is  the gauge invariant
object, i.e., different Nahm's data that correspond to the same spectral
curve describe the same monopole. It can be easily checked that the
spectral curve in the $A_2$ case does not depend on the functions
$\theta_i$. Therefore there is a freedom of fixing constants $c_i$ and
the function $\theta_1$ restricted by (\ref{gauge.a2}). It
is convenient to choose $c_1=c_2 =0$ and $\tan(\theta_1(s)) =2/{\cal
P}'(u)$. This choice leads to
\begin{equation}
e_1(s) = e_2 (s) = {{\cal
P}'(u)\over 2{\cal P}(u)} + i {1\over {\cal P}(u)},
\qquad e_3 (s) = - e_1^* (s).
\label{tetra}
\end{equation}
In the three-dimensional representation of $A_2$ given by
$$
E_{\alpha_1} = \pmatrix{0 &1&0 \cr 0&0&0 \cr 0&0&0}, \quad
E_{\alpha_2} = \pmatrix{0 &0&0 \cr 0&0&1 \cr 0&0&0}, \quad
E_{\alpha_3} = \pmatrix{0 &0&1 \cr 0&0&0 \cr 0&0&0},
$$
the Nahm data read
$$
{\bf T}  = \left(\pmatrix{0 &e_1&0 \cr -e_1^*&0&0 \cr 0&0&0},
\pmatrix{0 &0&0 \cr 0&0&e_1 \cr 0&-e_1^*&0},  \pmatrix{0 &0&-e_1^* \cr
0&0&0 \cr e_1&0&0} \right).
$$
With $e_1$ given by (\ref{tetra}) this is precisely the solution to
Nahm's equations with tetrahedral
symmetry found in \cite{HitMan:sym}.

As a second example we take the root system of $A_3$. There are six
positive roots $\alpha_i$, $i=1,\ldots 6$ which we choose so that
$\alpha_1,\alpha_2,\alpha_3$ are simple roots and $\alpha_4 = \alpha_1
+\alpha_2$, $\alpha_5 = \alpha_2+\alpha_3$, and $\alpha_6 =
\alpha_1+\alpha_2+\alpha_3$. Nahm's equations (\ref{nahm.a.1}) read
now
$$
{\bf e}_{\alpha_1}' = -{\bf e}_{\alpha_4} \wedge {\bf e}^*_{\alpha_2}
- {\bf e}_{\alpha_6} \wedge {\bf e}^*_{\alpha_5}, \qquad
{\bf e}_{\alpha_2}' = {\bf e}_{\alpha_4} \wedge {\bf e}^*_{\alpha_1}
- {\bf e}_{\alpha_5} \wedge {\bf e}^*_{\alpha_3},
$$
\begin{equation}
{\bf e}_{\alpha_3}' = {\bf e}_{\alpha_5} \wedge {\bf e}_{\alpha_2}
+ {\bf e}_{\alpha_6} \wedge {\bf e}^*_{\alpha_4}, \qquad
{\bf e}_{\alpha_4}' = {\bf e}_{\alpha_1} \wedge {\bf e}_{\alpha_2}
- {\bf e}_{\alpha_6} \wedge {\bf e}^*_{\alpha_3},
\label{nahm.a3}
\end{equation}
$$
{\bf e}_{\alpha_5}' = {\bf e}_{\alpha_2} \wedge {\bf e}_{\alpha_3}
+ {\bf e}_{\alpha_6} \wedge {\bf e}^*_{\alpha_1}, \qquad
{\bf e}_{\alpha_6}' = {\bf e}_{\alpha_1} \wedge {\bf e}_{\alpha_5}
- {\bf e}_{\alpha_3} \wedge {\bf e}_{\alpha_4}.
$$
The irreducibility conditions and the constraints (\ref{dep}) can be
easily solved to give
$$
{\bf v}_{\alpha_1} = {1\over\sqrt{2}}(0,1,1), \qquad {\bf
v}_{\alpha_2} = {1\over\sqrt{2}}(1,0,1), \qquad {\bf v}_{\alpha_3} =
{1\over\sqrt{2}}(0,-1,1),
$$
$$
{\bf v}_{\alpha_4} = {1\over\sqrt{2}}(-1,-1,0), \qquad {\bf
v}_{\alpha_5} = {1\over\sqrt{2}}(-1,1,0), \qquad {\bf v}_{\alpha_6} =
{1\over\sqrt{2}}(1,0,-1).
$$
These then are reflected by the relations between the components of
vector fields ${\bf e}_{\alpha_i}$, namely  ${e}_{\alpha_12}=
-{e}_{\alpha_13}$, ${e}_{\alpha_21}= -{e}_{\alpha_23}$, ${e}_{\alpha_32}=
{e}_{\alpha_33}$, ${e}_{\alpha_41}= -{e}_{\alpha_42}$, ${e}_{\alpha_51}=
{e}_{\alpha_52}$, ${e}_{\alpha_61}= {e}_{\alpha_63}$. Therefore there
are at most two different moduli of the components in each ${\bf
e}_{\alpha_i}$. Let $h_{\alpha_i, 1}$ be a modulus of the component of
${\bf e}_{\alpha_i}$ which occurs once and $h_{\alpha_i, 2}$ be a
modulus of the component of
${\bf e}_{\alpha_i}$ which occurs twice. The assumption that $g(s)$
does not depend on the phases of $e_{\alpha j}$
implies that for each $\alpha$ there is $k_\alpha\in \bf Z$
such that $\phi_{\alpha, i} - \phi_{\alpha, j} = (2k_\alpha
+1)\pi/2$, where $e_{\alpha i} = h_{\alpha, 1}\exp (i \phi_{\alpha, i})$
and $e_{\alpha j} = h_{\alpha, 2}\exp (i \phi_{\alpha, j})$.
Choosing $g(s)$ to be positive one finds
$$
f(s) = h_{\alpha ,1}(s)^2 + 2 h_{\alpha ,2}(s)^2, \qquad g(s) =
\sqrt{8}h_{\alpha ,1}(s) h_{\alpha ,2}(s) ,
$$
for all positive roots $\alpha$. Therefore there exist functions $u(s)$
and
$v(s)$, independent of $\alpha$,  such that $h_{\alpha 1} = u$ and
$h_{\alpha 2} = v$.

Next we take a closer look at the structure of phases
$\phi_{\alpha ,i}$. {}From (\ref{weier}) it follows that
$\phi_{\alpha, i} + \phi_{\beta, i} =\phi_{\alpha+\beta, i}
+k_{\alpha+\beta} \pi$ and that $k_{\alpha_6} =
k_{\alpha_4}-1$. Furthermore from (\ref{nahm.a3}) one deduces that the
phases $\phi_{\alpha ,i}$ are independent of $s$. Using
the gauge freedom we may choose $k_{\alpha_6} =0$. This choice reduces
the system of equations (\ref{nahm.a3}) to the two-dimensional problem
\begin{equation}
{d v\over ds} = 2uv ,\qquad  {d u\over ds} = 2(v^2 -u^2),
\label{nahm.a3.1}
\end{equation}
a solution of which is given by
\begin{equation}
v(s) = c^{1/4} {{\cal P}(t) +i\over {\cal P}(t) -i}, \qquad u(s) =
{1\over 2} {d\over ds} \log v(s),
\label{a3.sol}
\end{equation}
where $t = \pm \sqrt{2} c^{1/4} e^{i\pi/4} s$, $c = v^2(v^2 - 2u^2)$
is the integration constant and ${\cal P}$ is the Weierstrass elliptic
function given by the equation ${\cal P}'(t)^2 = {\cal P}(t)({\cal
P}(t)^2 -1)$. The solution (\ref{a3.sol}) is equivalent to the
octahedral solution to Nahm's equations found in
\cite{HitMan:sym}.\bigskip

{\bf 4.} In  this brief paper we have described a method of deriving
symmetric Nahm's data from the root systems of simple Lie algebras of
$A_n$ type. We
have shown that the tetrahedral and octahedral monopole configurations
correspond to root systems of $A_2$ and $A_3$ type. For the sake of
brevity and clarity we
skipped all the proofs and detailed derivations of the results. We
intend to present them in a forthcoming full-size article. We also
intend to give new solutions to Nahm's equations as well as proofs of
non-existence of solutions for certain root systems (such as $A_4$ for
example). Finally we would like to mention that each solution to
Nahm's equation obtained in the way described in this paper gives
rise to  an integrable model of particles interacting on the
line. A detailed description and solutions of these models are
currently being investigated.\bigskip

{ACKNOWLEDGEMENTS.} We are indebted to Nick Manton for
encouragement and many interesting discussions.

\end{document}